\documentclass[a4paper,twoside]{article}

\usepackage{epsfig}
\usepackage{subfigure}
\usepackage{calc}
\usepackage{amssymb}
\usepackage{amstext} 
\usepackage{amsthm} 
\usepackage{xspace}
\usepackage{focus}
\usepackage[table,xcdraw]{xcolor}  
\usepackage{textcomp}
\usepackage{url}
\usepackage{array,multirow,graphicx}

\usepackage{SCITEPRESS}      

\begin{document}

\newcommand{\focust}{\textsc{Focus}$^{ST}$}
\newcommand{\focuse}{\textsc{Focus}$^{E}$}

\title{(Auto)Focus approaches and their applications:\\
A systematic review}


\author{\authorname{Maria Spichkova}
\affiliation{School of Science, RMIT University, 414-418 Swanston Street, 3001, Melbourne, Australia} 
 \email{maria.spichkova@rmit.edu.au}
}

\abstract{
\Focus,
a framework for formal specification and development of interactive systems, 
was introduced approx. 25 years ago.
Since then this approach was broadly used in academic and industrial studies, as well as provided a basis  for a number of another frameworks focusing on particular domains, and  
for the AF3 modelling tool. 
In this paper we provide a literature review of the corresponding approaches, academic case studies and industrial applications of these methods.
}

\onecolumn 
\maketitle 
\normalsize \vfill

\section{\uppercase{Introduction}}

\Focus,
a framework for formal specification and development of distributed interactive systems, 
provides models and formalisms for a step-wise specification and development. 
\Focus was introduced by Broy et al. \cite{broy1992design} approx. 25 years ago. Since then, the framework was extended and gave a basis for a number of other approaches. Many case studies from several application domains were conducted, within both academic and industrial projects. 
One of the most comprehensive description of the formal background of the framework was presented by Broy and St{\o}len \cite{focus}, 
and became a highly cited publication (more than 540 citations\footnote{According to the Google Scholar, retr. 20/11/17}). 
The goal of this approach is to support a modular system development with the help of component specifications, having precisely identified interfaces and formal refinement concepts. 
Focus is based on  a clear mathematical theory and aims on system development carried out in a systematic way. 

The tool-support for this approach is provided by the AutoFocus CASE tool.
The prototype was  initially presented by Broy et al. \cite{broy1999autof}, and later extended and refined by two next generations of the tool. 
The third  generation of AutoFocus, AutoFocus3 or AF3,
is a powerful open-source  tool.  
It's emphasis is on the development of embedded systems using models from the requirements to the hardware architecture. 
The latest version of AF3 provides advanced features to support the user: 
formal analyses, synthesis methods, space exploration visualization, etc.

The research on the \Focus-related approaches is an active topic. 
The most resent related works were presented by Alzahrani et al.
\cite{alzahrani2017temporal}, where aims was to apply property-based testing on formal models with temporal properties, and by Kanav and Aravantinos
\cite{AF3modevva2017}, introducing the modular transformation from AF3 to the \emph{nuXmv} symbolic model checker.
As over the last 25 years many theoretical and applies work was conducted within the \Focus approach, an overview of existing work is required.  
In this paper, we are going to provide a systematic review of the corresponding approaches as well as on the case studies they were applied on.

\emph{Outline:} 
Section \ref{sec:focus} provided a high-level overview of the \Focus ideas. 
Spatio-Temporal view on modelling of interactive systems is presented in Section \ref{sec:focust}.
Section \ref{sec:auto} discusses the tool support for the approach.
Section \ref{sec:methodologies} introduces methodological extensions 
of the \Focus approach as well the frameworks built on its basis.  
Section \ref{sec:applications} reviews the academic and industrial case studies conducted using \Focus or related approaches.
Section \ref{sec:hofm} discusses human-oriented aspects of the related formal methods.
Finally, Section \ref{sec:conclusions} summarises the paper.

\section{\uppercase{Focus}}
\label{sec:focus}
 
A distributed system in \Focus is specified by its logical \emph{components} 
 connected via \emph{channels}, which  are  directed and order preserving. 
Components  work independently of each other or interact, exchanging
  information in terms of \emph{messages} of specified types.
  
The formal meaning of a \Focus specification is an \emph{input/output relation}, i.e. 
 the relation between the
communication histories for the external \emph{input} and \emph{output channels}. 
The specifications can be 
\begin{itemize}
\item structured into a number of
formulas each characterizing a different kind of property;
\item elementary or composite: 
  \begin{itemize}
  \item Elementary specifications can be
untimed, timed, and time-synchronous;
  \item Composite specifications  are
built hierarchically from the elementary ones.
  \end{itemize}
\end{itemize}
The central concept in \Focus are \emph{streams}, that represent 
communication histories of \emph{directed channels}. 
Streams in \Focus are functions mapping the indexes in their domains
 to their messages. 
For any set of messages $M$, $M^\omega$ denotes the set of all streams,
$M^\infty$ and $M^*$ denote  the sets of all infinite and all finite
streams respectively.
 $M^{\underline{\omega}}$ denotes the set of all 
timed streams, $M^{\underline{\infty}}$ and $M^{\underline{*}}$ denote the sets of 
all infinite and all finite timed streams respectively.

\[ \begin{array}{l}
M^\omega \stackrel{\mathrm{def}}{=} M^* \cup M^\infty 
\\
M^* \stackrel{\mathrm{def}}{=} {\bigcup}_{n \in \mathbb{N}}([1..n]\to M)
\\
M^\infty \stackrel{\mathrm{def}}{=} \mathbb{N}_{+} \to M 
\end{array}
\]
 Thus,  \emph{timed stream} is represented by a sequence of messages and {\it time ticks}, 
the messages are also listed in their order of transmission. The ticks model a discrete notion of time.  
\[ 
\begin{array}{l}
M^{\underline{\omega}}$ =  $M^{\underline{*}} \cup M^{\underline{\infty}}\\
M^{\underline{*}} \stackrel{\mathrm{def}}{=} 
\bigcup_{n \in \mathbb{N}}([1..n]\to M \cup \{ \surd \})\\
M^{\underline{\infty}} \stackrel{\mathrm{def}}{=} 
\mathbb{N}_+ \to M \cup \{ \surd \}
\end{array}
\] 

\section{\uppercase{Spatio-Temporal View: Focus$^{ST}$}}
\label{sec:focust}

Timing aspects of \Focus as well as the corresponding optimisations of the specification layout were discussed in \cite{spichkova2012time}, which was the first step towards elaboration the \focust framework.

In both frameworks, specifications are based on the notion of \emph{streams}.
However, in the original \Focus  input and output streams of a component are mappings
 of natural numbers $\Nat$ to single messages,
whereas a \focust\  stream %
 is a mapping from $\Nat$ to lists of messages  within the corresponding time intervals.
 Moreover, the syntax of \focust\ is particularly devoted to specify spatial (S) and timing (T) aspects in a comprehensible fashion,
which is the reason to extend the name of the language by $^{ST}$.

The \focust\ specification layout \cite{spichkova2016spatio,spichkova2014modeling} is similar to \Focus (which layout was inspired by Z specification language, cf. \cite{Spivey_88,Spivey_92}), 
but it has many new features to increase the readability and understandability of the specification.  
The \focust\ specification layout 
is based on human factor analysis within formal methods (see Section~\ref{sec:hofm}). 

In \focust specifications, input and output streams of a component are always timed, as spatio-temporal aspects are the core of the framework. 
The (timed) streams are mappings from $\Nat$ to lists of messages  within the corresponding time intervals. 
Thus, these streams are infinite per default, but they could be empty completely or from a certain point which is represented by empty time intervals $\nempty$.  
More precisely, \focust has streams of two kinds:
	\begin{itemize}
	\item
	\emph{Infinite timed streams} (denoted by $M^{\underline{\infty}}$) are used to represent the input and the output streams;
	\item
	\emph{finite timed streams} (denoted by $M^{\underline{*}}$) are used  
	to argue about a timed stream that was truncated 
	at some point of time.
	\end{itemize}  
 Infinite timed streams of type $T$ are defined  by a functional type 
\[
\Nat\ \to \nfst{T}
\] 
Finite timed streams of type  $T$  are defined by list of lists over this type, i.e., 
\[
\nfst{(\nfst{T})}
\]
where $\nfst{T}$ denotes a list of elements of type $T$.

The \focust ideas were applied within the approach on an intelligent route planning for a public transport system within a sustainable Smart City, cf.  \cite{spichkova2015trade,spichkova2015route}.
 
Spatio-temporal models for formal analysis and property-based testing were presented in 
\cite{alzahrani2016spatio,alzahrani2017temporal} by Alzahrani et al.
The authors aimed to to apply property-based testing on \focust and TLA models  with temporal properties.

\section{\uppercase{AutoFocus}}
\label{sec:auto}

AutoFocus tool was developed based on the \Focus theory. 
The first two prototype versions were replaced by AF3 (AutoFocus 3, cf. \cite{holzl201013}) tool that provides a various functionalities and supports several development phases.\footnote{\url{https://af3.fortiss.org/}} 
AF3 embeds the core modelling artefacts, is open source, and has a well defined formal syntax behind all its modelling elements.
Source code of AutoFocus3 models are coded in XML, which makes it easy to parse and to analyse.   
The tool was applied as a part of tool chain within a number of development methodologies, cf. Section  \ref{sec:methodologies}.

H{\"o}lzl et al. proposed an idea of AutoFocus tool chain and its application for the development of safety-critical systems \cite{holzl2010autofocus,holzl2010safety}. 
An extended and refined version of the approach was later presented in  \cite{spichkova2012verified}. 
This approach  also was elaborated using AutoFocus 2 and allows a direct application within AF3.
 
Campetelli et al. \cite{campetelli2011user} introduced an approach to a user-friendly model
checking integration in model-based development using AF3. 
The approach supports support
two different model checkers for the model and
implementation code verification: SMV and TVARC.

Kondeva et al. 
\cite{kondeva2013seamless} presented how the integrated system views on several levels of abstraction are implemented in AF3, to allow the development of embedded systems.

Teufl et al. \cite{teufl2013mira} presented an integrated with AF3 tooling-framework to experiment with model-based Requirements engineering.

Cimatti  and Tonetta \cite{cimatti2016temporal} introduced 
a temporal logics approach to contract-based design, integrated within AF3. 

AF3 tooling concepts for  
model-based development of embedded systems was presented in~\cite{aravantinos2015autofocus}.

A number of approaches on scheduling and deployment were also integrated within  AF3.
Voss and Sch{\"a}tz  presented an approach on scheduling shared memory multicore architectures in AF3 using SMT solvers \cite{voss2012scheduling}, as well as an approach
on deployment and scheduling Synthesis for mixed-critical shared-memory applications \cite{voss2013deployment}.

An approach on generating formal specifications from system AF3 models was presented in \cite{spichkova2013we}. 
Applying this approach would allow to solve  the problem with outdated system documentation  by making the documentation updates automatically: 
an up-to-date formal specification could be generated from the model if the model is frequently changed.  
The next step in this detection was presented in \cite{vo2016model}: this approach  allowed  generation of natural language specifications from AF3 models.

\section{\uppercase{Methodologies}}
\label{sec:methodologies}

A number of software and system development methodologies were introduced involving   \Focus or the framework created in its basis, as well as involving  an extensive use of AutoFocus tools.

One of the core features of the current \Focus framework, the specification of the black box behaviour of data flow components by characterizing the
relation between the input and the output histories, was initially introduced in \cite{broy1994specification}. The authors distinguished between three
main specification classes: time-independent specifications, weakly
time-dependent and strongly time-dependent specifications. 
Data flow components were formally specified by sets of timed stream processing functions.
Specifications describe such sets by logical formulas. The proposed solution allowed to handle the Brock/Ackermann anomaly~\cite{brock1981scenarios}.  
A further analysis of considering a component as a black box that
is a physical encapsulation of related services was presented in \cite{broy1997component}.
Philipps and Rumpe proposed a stepwise refinement of data flow architectures, cf. \cite{philipps2014stepwise}.

Another core feature of the current \Focus framework,  a method for the specification of reactive asynchronous components with a concurrent access interface, was discussed in
\cite{broy1995advanced}.

The approach for structured specifications and implementation of non-deterministic data types
was introduced in \cite{walicki1995structured}.   

St{\o}len used relations on \Focus streams to solve the RPC-memory specification problem, 
cf.~\cite{stolen1996using}.
 
Broy applied the \Focus-based theory to  introduce a logical basis for component-based system engineering \cite{Broy99alogical} as well as for 
specification of interface behaviour of multifunctional systems
\cite{Broy2010MSS}, i.e.,    systems that offer a variety of functions for different purposes and use cases. 
In \cite{Broy2010MSS}, service hierarchies specify multifunctional systems in terms of 
services (provided sub-functions) taking into account their mutual relationships and dependencies. Each service is specified independently and the specification is added to the service hierarchy, which then describes the functionality of multifunctional systems.

The Janus approach \cite{broy_janus,Broy2007services} was introduced 
to formally design of services and layered architectures, based 
 on the \Focus theory of distributed systems. 
 A Janus service, like a \Focus component, has a syntactic interface, but, 
 in comparison to a component, a service has a partial behaviour.

A methodology for modularised specification and verification of distributed time-triggered systems was proposed in \cite{botaschanjan2005towards,botaschanjan2006towards}. 
The authors applied \Focus to specify the systems formally. 

A number of \Focus-related works were focusing on refinement aspects:
from a general analysis of compositional refinement of interactive systems \cite{broy_refinement}  as well as the notions of abstraction and causality on the specification level  \cite{broy_time_abstr} to the refinement-based verification of interactive systems \cite{spichkova2008refinement,spichkova2009refinement,botaschanjan2008correctness} and its relation to the architectural aspects of software and system development \cite{spichkova2011refinement}. 
The corresponding methodology of architectural decomposition  was discussed in
\cite{spichkova2010architecture,spichkova2011decomp}. 
Transformation of semi-formal requirements to formal \Focus specifications  was proposed in \cite{spichkova2010semiformal}.

A specification and proof methodology  ``\Focus on Isabelle''
 \cite{spichkova2013we,spichkova,spichkova2008focus}. 
supports an alignment on the future proofs during specification phase
to make the proofs simpler and appropriate for application not only in theory but also in practice.  
Given a system represented in \Focus or \focust, the methodology
allows us to verify system properties by translating the specification to a Higher-Order Logic and 
subsequently using an interactive semi-automatic theorem prover~\cite{npw}. 
or the point of disagreement will be found. 
Another advantage of the methodology is a well-developed theory of composition, both for general 
and cryptographic properties \cite{spichkova2008formal,spichkova2012component}. 
 
Feilkas et al. presented  a methodology for the top-down 
development of automotive software systems \cite{feilkas2009top,feilkas2011refined}. 
The core artefact applied within the methodology was AutoFocus tool and the corresponding extension for test case generation. The methodology was elaborated using AutoFocus 2, however, it can be applied without any changes for AF3.
An extension of this \Focus-based methodology to the domain of cyber-physical systems was proposed in \cite{spichkova2012cyber}.

Vogelsang et al. \cite{vogelsang2014supporting} proposed a model-based approach
that starts from informal use cases and enables
a stepwise formalization of functional requirements 
to be linked to the architecture of the
system.  The approach utiled AF3 as the modelling framework.  

An approach introduced by Doby et al. \cite{dobi2015model} utilized \Focus to provide 
an efficient hazard and impact analysis for automotive mechatronics systems. 

Another approach based on \focust, allows analysis of component dependencies    \cite{spichkova2014formalisation}. This was later extended to framework for formal analysis of dependencies among services \cite{spichkova2014towards}.

The theory of processes extending the \Focus framework was introduced in 
\cite{spichkova2011focus}. 
This theory was further extended to have
a formal model of processes that is compatible with the component/ data flow view, cf. 
\cite{spichkova2015formal,mise2015process}. In many cases, it is beneficial to specify on the same abstraction level not only system components  
but also processes within the system, however, if we have to apply different frameworks to analyse both views,  
the system model becomes hard to read and to understand. 
The presented in \cite{mise2015process} approach provides 
a solution how to cover this gap and to  reconcile   component and process  views.

\section{\uppercase{Case Studies}}
\label{sec:applications}

The summaries of the first case studies in \Focus was presented approx. 25 years ago, in 1992 and 1994, cf. \cite{broy1992summary,broy1994summary}. 
The second summary was presented 5 years later, in 1997 \cite{broy1997summary}.

Specification and refinement of a buffer of length one was presented in \cite{broy1995buffer}.
The first version of the \Focus specification for a steam boiler system was introduced in \cite{broy1999boiler}.

A Trading System case study was used as common example 
for modelling approaches of component-based systems within the CoCoME contest to compare software component models. 
Broy at al. presented within this contest a systematic model-based approach of the engineering of distributed systems, which was based on the application of \Focus and AutoFocus2, cf. \cite{cocome}. The authors provide tool support around the AutoFocus2 tool,
that enables us to execute our specified models in a distributed environment
targeted to the CoCoME example.

Formal \Focus specifications of FlexRay and FTCom were introduced in \cite{kuhnel2007fault,kuhnel2006flexray,kuhnel2006upcoming}, where  
FlexRay is a time-triggered communication protocol,  and 
FTCom is the communication layer of a time-triggered 
operating system OSEKtime.
An operating system OSEKtime  
was developed by the European Automotive Consortium OSEK/VDX in accordance to the time-triggered paradigm. 
OSEK is a standards body, founded by German automotive company consortium,  
which included many industrial partners (e.g., BMW, Bosch, Siemens, etc.) as well as the University of Karlsruhe. 
The French automotive manufacturers Renault and PSA Peugeot Citroen had a similar consortium, VDX.
In 1994, a new consortium OSEK/VDX\footnote{\url{http://www.osek-vdx.org}} was created, based on OSEK and VDX.
FlexRay and FTCom were introduced for 
the fault-tolerant communication   by the FlexRay Consortium and
OSEK/VDX respectively.
Authors presented \Focus specifications of FlexRay and FTCom that allow us to argue about
the properties of FlexRay and FTCom in a formal manner. 
The \Focus specification of FlexRay was also verified methodology  ``\Focus on Isabelle'', discussed in Section \ref{sec:methodologies}, confirming that this 
specification of FlexRay fulfils the FlexRay requirements, cf. \cite{spichkova2006flexray}.
A number of case studies on the modelling of autonomous systems were presented in \cite{spichkova2015towards,spichkova2017autonomous,spichkova2016automotive}.

Within the methodology  ``\Focus on Isabelle'', three case studies were elaborated using the version of Isabelle framework of 2007:
\begin{itemize} 
\item data transmission (FlexRay communication protocol), 
\item process control (Steam Boiler System),  
\item memory and processing components (Automotive-Gateway System). 
\end{itemize}
These case studies were later conducted using the Isabelle  version of 2012, 
which allows to use the Isar language \cite{Isar} providing human-readable proofs in HOL), cf. \cite{FocusStreamsCaseStudies-AFP}. 
A further optimisation of the case studies on the verification level was proposed in \cite{spichkova2017human}: the authors introduced a human-oriented methodology for analysis of the dependencies between lemmas within the provided set of proofs, extending the ``\Focus on Isabelle'' methodology.

The case study introduced in \cite{CryptoBasedCompositionalProperties-AFP}, presents 
a \Focus formalisation of the security property of data secrecy along with  
the corresponding definitions and   Isabelle/HOL proofs.

B{\"o}hm et al. \cite{bohm2014formal} 
reported  successful results of a project conducted in collaboration between Siemens AG, fortiss GmbH and TU Munich. The goal was to evaluate SPES modeling framework (SPES MF)  implemented within AF3. B{\"o}hm et al. performed a case study, on modelling of requirements and functionality for a part of a Siemens train automation system. The results demonstrated advantages of application the SPES MF and AF3 within this context.

Campetelli et al. \cite{campetelli2015model} presented an industrial case study from the automation domain,
focusing on the control software components: 
a case example of a seawater desalination plant was
modelled in AF3 according to the proposed SPES development method.
 
Spichkova et al. \cite{spichkova2016formal} illustrated using formal models for intellingent speed validation and adaptation.
Formal specification of Chiminey platform \cite{Chiminey,yusuf2017chiminey}, which provides a reliable computing and data management service, as well as its refinements and extensions were presented in 
\cite{ChimineyICPADS,spichkova2016managing,spichkova2017towards}. 

Zamansky et. al. \cite{zamansky2016formal} reviewing some recent large-scale industrial projects in which formal methods (including \Focus and AutoFocus) have been successfully applied. The authors also covered some aspects of teaching formal methods for software engineering, including \Focus and AutoFocus, cf. \cite{spichkova2016teaching,simic2016enhancing}.

\section{\uppercase{Human-Oriented Aspects}}
\label{sec:hofm}

Sch{\"a}tz et al. \cite{schatz1996graphical} 
argued almost 20 years ago that that formal techniques are indeed useful for practical application, but they should be put to indirect use. 
To demonstrate this approach, the authors analysed   two pragmatic graphical description techniques, taken from the field of telecommunication.
 The analysis was  targeting on the information content of the techniques and their application in the process of specification development. The authors defined the techniques  formally, and introduced based on these formal definitions a number of  development steps and their graphical counterparts. This work can be seen as the first step towards graphical specification style within the \Focus framework as well as to the  AutoFocus tool.

An approach presented in \cite{hffm_spichkova,spichkova2013design}
aims to apply the engineering psychology achievements to the design of formal methods, focusing on the specification phase of a system development process.
Its core ideas originated from the analysis of the \Focus framework and also led to 
an extended version of the framework, \focust.  

\cite{spichkova2015human} introduced an research on incorporation of the human factors engineering into the software development process:  The authors proposed to apply the 
human factors analysis not only the level of requirements specification and formal system modelling, but also to guide various testing tasks.


\section{\uppercase{Conclusions}}
\label{sec:conclusions}

This paper provides a literature review of the methodological approaches, academic case studies and industrial applications of 
\Focus and \focust, 
the framework for formal specification and development of interactive systems, as well as
the AF3 (AutoFocus 3) modelling tool developed based on the  \Focus theory.

\Focus was introduced approx. 25 years ago, and extensively used since then in many academic and industrial projects.  
The literature review covers more than 80 publications on \Focus, \focust  and AutoFocus research, from 1992 when the first publication on \Focus appeared till 2017.

\bibliographystyle{abbrv} 

\vfill
 
\end{document}